\documentclass[12pt,]{article}
\usepackage{graphicx}
\usepackage{amssymb}
\begin{document}

\begin{flushleft}
\noindent  {\Large{\bf Algorithm for Probing the Unitarity\\ of
Topologically Massive Models}}
\end{flushleft}
 \vskip 1cm

\begin{minipage}[b]{13.2cm}\parbox[t]{1.2cm}{}\hfill\parbox[t]{12.2cm}{
{\bf Antonio Accioly$^{1,\;2}$ and Marco Dias$^1$}

\vskip 1cm

\rule{0.92\textwidth}{0.01mm}
 \small{ \noindent An uncomplicated
and easily handling prescription that converts the task of
checking the unitarity of massive, topologically massive, models
into a straightforward algebraic exercise, is developed. The
algorithm is used to test the unitarity of both topologically
massive higher-derivative electromagnetism and topologically
massive higher-derivative gravity. The novel and amazing features
of these effective field models are also discussed. }
\rule{0.92\textwidth}{0.01mm} \vskip 0.3cm


\noindent {\small {\bf PACS NUMBERS:}} 11.10.Kk; 04.60.Kz;
11.10.St. }
\end{minipage}

\vskip .5cm

\noindent {\bf 1. INTRODUCTION}

\vskip .5cm

The momentous discovery that there are dynamics possible for gauge
theories in an odd number of space-time dimension that are not
open to those in an even number, allowed the construction of field
models endowed with novel and amazing properties. In
three-dimensions, for instance, the addition of a topologically
massive Chern-Simons term to the fundamental Lagrangian for a
gauge-field gives rise to a gauge-invariant theory (Deser {\it et
al}, 1988a,b). Indeed, this term has a coupling that scales like a
mass, but unlike the ways in which gauge fields are usually given
a mass, no gauge symmetry is broken, although parity is. Of
course, the addition of the esoteric Chern-Simons term is
certainly not the unique mass-generating mechanism for gauge
fields. We can also utilize for this purpose the well-known
Proca/Fierz-Pauli, or the more sophisticated higher-derivative
electromagnetic/higher-derivative grav-

 \vskip .3cm
 {\small {\noindent $^1$Instituto de F\'{\i}sica Te\'orica,
 Universidade Estadual Paulista, S\~ao Paulo, Brazil.

\noindent $^2$To whom correspondence should be addressed at
Instituto de F\'{\i}sica Te\'orica,
 Universidade Estadual Paulista,  Rua Pamplona 145, 01405-000 S\~ao Paulo, SP,
 Brazil; e-mail:accioly@ift.unesp.br. }}

\newpage \noindent itational, terms. In this vein, it would be
interesting to analyze the new physics that emerges from the
models obtained by enlarging Maxwell (Einstein)-Chern-Simons
theory through the Proca (Fierz-Pauli), or higher-derivative
electromagnetic (gravitational), terms. Our aim here is to study
the three-term models with higher-derivatives. Interesting enough,
these models are gauge-invariant; besides, they possess rather
unusual and exciting properties. In fact, as we shall see,  in the
context of the electromagnetic models, an attractive interaction
between equal charge scalar bosons can occur which leads to an
amazing   planar electrodynamics: scalar pairs can condense into
bound states; while in the framework of the gravitational systems,
unlike what happens within the context of the odorless and insipid
three-dimensional general relativity, there exists both attractive
and repulsive gravity. We can also have a null gravitational
interaction, such as in three-dimensional gravity that is trivial
outside the sources.

We present an algorithm for probing the unitarity of massive ,
topologically massive, models (MTM) in Section 2, which is quite
simple to use. This procedure converts the hard task of checking
the unitary of the MTM in a trivial algebraic exercise. It is
utilized  to test the unitarity of both topologically massive
higher-derivative electromagnetism (TMHDE) and topologically
massive higher-derivative gravity(TMHDG), in Section 3.  The novel
and amazing features of the electromagnetic models are discussed
in Section 4, while those of the gravitational ones are analyzed
in Section 5. We conclude in Section 6 with some discussions and
comments. We use natural units throughout. \vskip .5cm

\noindent {\bf 2. ALGORITHM FOR PROBING THE UNITARITY OF MASSIVE,
TOPOLOGICALLY MASSIVE , MODELS} \vskip .5cm

 To probe the tree
unitarity of the massive, topologically massive, models, we  will
make use of the procedure that consists basically in saturating
the propagator with external conserved currents, compatible with
the symmetries of the system. The unitarity of the models depends
on the sign of the residues of the saturated propagator
($SP$)---the  unitarity is ensured if the residue at each simple
pole of the $SP$ is positive (propagating modes) or zero
(non-propagating modes). Note that we are using the loose
expression ``the residue's sign is equal to zero" as synonymous
with ``the residue is equal to zero".

The idea here is to construct a simple algorithm for analyzing the
unitarity  of the massive, topologically massive, models, using
the procedure we have just outlined. We begin by building the
prescription for the massive, topologically massive,
electromagnetic  models (MTME); next we construct the algorithm
for the massive, topologically massive,
 gravitational models (MTMG). \vskip .5cm

\noindent {\it 2.1 Algorithm for analyzing the unitarity of the
MTME} \vskip .5cm

 The saturated propagator related to the MTME, can be written as

\begin{eqnarray}
SP_{\mathrm{MTME}} = J^\mu \left( O^{-1}_{\mathrm{MTME}}
\right)_{\mu\nu} J^\nu,
\end{eqnarray}

\noindent where $J$ and $O^{-1}$ are, respectively, the conserved
current and the propagator concerning the specific massive,
topologically massive, electromagnetic model which we are
interested in probing the unitarity. Our next step is to obtain
the propagator associated with the model at hand. Consider, in
this direction, the Lagrangian for the MTME, namely ${\mathcal
L_{\mathrm {MTME}}} = {\mathcal L_{\mathrm{E}}} + \epsilon
{\mathcal L_{\mathrm{gf}}} + {\mathcal L_{\mathrm{T}}}$, where
${\mathcal L_{\mathrm{E}}}$ is the Lagrangian associated with the
electromagnetic part of the model, ${\mathcal L_{\mathrm{gf}}}$ is
a gauge-fixing Lagrangian, $\epsilon$ is a parameter equal to
$+1$,
 if ${\mathcal L_{\mathrm{E}}}$ is gauge-invariant, or $0$, if ${\mathcal
L_{\mathrm{E}}}$ is not gauge-invariant, and ${\mathcal
L_{\mathrm{T}}} \equiv \frac{s}{2} \varepsilon_{\mu\nu\rho} A^\mu
\partial^\nu A^\rho$ is the Chern-Simons term, with $A^\mu$ being
the three-dimensional vector potential and $s>0$ the topological
mass. This Lagrangian, of course, can be written as ${\mathcal
L_{\mathrm{MTME}}} = \frac{1}{2} A^\mu O_{\mu\nu} A^\nu$. Now, it
is important for the success of the method that we can  find  a
basis for expanding the wave operator and, consequently, the
propagator, such that when one contracts their  basis vectors with
$JJ$, the greatest possible number of cancellations may be
obtained. The basis $ \{ \theta, \omega, S\}$, for instance, where
$\theta_{\mu\nu} \equiv \eta_{\mu\nu} - \frac{\partial_\mu
\partial_\nu}{\Box}$ and $\omega_{\mu\nu} = \frac{\partial_\mu
\partial_\nu}{\Box}$ are, respectively, the usual transverse and
longitudinal vector projector operators, $S_{\mu\nu} \equiv
\varepsilon_{\mu\rho\nu}  \partial^\rho$ is the operator
associated with the topological term, and $\eta_{\mu\nu}$ is the
Minkowski metric, does the job since $J\omega J= JSJ =0$. The
algebra obeyed by these operators is displayed in Table 1. Our
signature conventions are $(+,-,-)$,
$\varepsilon^{012}=+1=\varepsilon_{012}$.

Expanding $O$ in the basis $\{\theta,\omega,S \}$, yields $ O =
a\theta + b\omega + cS $. With the help of Table 1, we promptly
obtain
\begin{equation}
O^{-1}_{\mathrm{MTME}} = \frac{a}{a^2 + c^2 \Box} \theta +
\frac{1}{b} \omega - \frac{c}{a^2 + c^2 \Box} S.
\end{equation}

Inserting eq.2 into eq.1, we get

\begin{eqnarray}
SP_{\mathrm{MTME}} =\frac{a}{a^2 + c^2 \Box} J^\mu J_\mu.
\end{eqnarray}

\noindent Note  that only  the $\theta$-component of
$O^{-1}_{\mathrm{MTME}}$ contributes to the calculation of
$SP_{\mathrm{MTME}}$ .

\begin{table}[h]
\caption{Multiplicative table for the operators $\theta, \;\omega$
and $S$. The operators are supposed to be in the ordering ``row
times column".} \vskip .5cm
\begin{center}
\begin{tabular}{cccc}
\hline & $\theta$ &$\omega$&$S$\\\hline $\theta$& $\theta$& 0
&$S$\\ $\omega$& 0 &$\omega$&0\\ $S$&$S$&0&$-\Box\theta$ \\ \hline
\end{tabular}
\end{center}
\end{table}

Before going on, we need a lemma. \vskip .5cm

{\bf Lemma 1.} {\it If $m \geq 0$ is the mass of a generic
physical particle associated with the MTME and $k$ is the
corresponding momentum exchanged, then $J_\mu J^\mu|_{k^2= m^2} <
0$.}

\vskip .5cm

{\bf Proof.} To begin with, let us expand the current in a
suitable basis. The set of independent vectors in momentum space,
\begin{equation}
k^\mu \equiv \left(k^0, {\bf k}\right)\;,\;{\tilde{k}}^\mu \equiv
\left(k^0, -{\bf k}\right)\;,\;\varepsilon^\mu \equiv
\left(0,\;{\vec{\epsilon}}\;\right),
\end{equation}
\noindent where ${\vec{ \epsilon}}$ is a unit vector orthogonal to
${\bf k}$, serves our purpose. Using this basis, $J^\mu (k)$ takes
the form

$$J^{\mu} = A k^\mu  + B{\tilde k}^\mu  + C \varepsilon^\mu . $$

On the other hand, the current conservation gives the constraint
$A (k_0^2 - {\bf k^2})- B (k_0^2 + {\bf k^2}) = 0$, which allows
to conclude that $A^2 > B^2$. Now, it is trivial to see that
$J_\mu J^\mu = k^2 (B^2 - A^2) -C^2$. Consequently, $J_\mu
J^\mu|_{k^2=m^2} < 0$.

\vskip .5cm

We are now ready to present the algorithm for probing the
unitarity of the MTME. \vskip .5cm

{\bf Algorithm 1.} {\it Calculate the $\theta$-component of the
propagator in the basis $ \{ \theta, \omega, S\}$ which, for
short, we shall designate as $f_\theta$. Next, determine the signs
of the residues at each simple pole of $f_\theta$. If all the
signs are $\leq 0$, the model is  unitary; if at least one of the
signs is positive, the system is non-unitary .} \vskip .5cm

\noindent {\it 2.2 Algorithm for  analyzing the unitarity of the
MTMG} \vskip .5cm The Lagrangian  for the MTMG can be written as
${\mathcal L_{\mathrm {MTMG}}} = {\mathcal L_{\mathrm{G}}} +
\epsilon {\mathcal L_{\mathrm{gf}}} + {\mathcal L_{\mathrm{T}}}$,
where ${\mathcal L_{\mathrm{G}}}$ is the Lagrangian concerning the
gravitational part of the theory, and ${\mathcal L}_{\mathrm T}
\equiv \frac{1}{\mu} \varepsilon^{\lambda\mu\nu}
{\Gamma^\rho}_{\lambda\sigma}  \left(
\partial_\mu {\Gamma^\sigma}_{\rho\nu} +
\frac{2}{3}{\Gamma^\sigma}_{\mu\beta} {\Gamma^\beta}_{\nu\rho}
\right)$ is the Chern-Simons Lagrangian, with $\mu>0$ being a
dimensionless parameter, whereas the  corresponding $SP$  is given
by

\begin{equation}
SP_{\mathrm{MTMG}} = T^{\mu\nu} \left( O^{-1}_{\mathrm{MTMG}}
\right)_{\mu\nu,\;\rho\sigma} T^{\rho\sigma},
\end{equation}
\noindent where $T^{\mu\nu}$ is the conserved current which,
obviously, is symmetric in the indices $\mu$ and $\nu$.  Our
conventions are ${R^\alpha}_{\beta\gamma\delta} = -
\partial_\delta {\Gamma^\alpha}_{\beta\gamma} + ... \;, R_{\mu\nu}
= {R^\alpha}_{\mu\nu\alpha}, R= g^{\mu\nu} R_{\mu\nu}$, where
$g_{\mu\nu}$ is the metric tensor, and signature $(+, -, -)$. To
calculate the $SP_{\mathrm{MTMG}}$, we need  to know the
propagator beforehand . This can be done by linearizing
 ${\mathcal L_{\mathrm {MTMG}}}$ . Setting $ g_{\mu\nu} =
\eta_{\mu\nu} + \kappa h_{\mu\nu}$, where $\kappa$ is a constant
that in four dimensions is equal to $\sqrt {32 \pi G}$, with $G$
being Newton's constant, we can rewrite  the linearized Larangian
as ${\mathcal L}_{\mathrm {MTMG}}^{\mathrm (lin)} = \frac{1}{2}
h_{\mu\nu} O^{\mu\nu,\;\rho\sigma} h_{\rho\sigma}$. It is
extremely convenient to expand $O$ in the basis $\{P^1, P^2, P^0,
{\overline P}^{\;0}, {\overline{\overline P}}^{\;0} , P \}$, where
$ P^1, P^2, P^0, {\overline P}^{\;0},$ and ${\overline{\overline
P}}^{\;0}$, are the usual three-dimensional Barnes-Rivers
operators (Rivers, 1964; Nieuwenhuizen, 1973; Stelle, 1977;
Antoniadis and Tomboulis, 1986), namely,

\begin{eqnarray}
P^1_{\mu\nu,\;\rho\sigma} = \frac{1}{2} \left( \theta_{\mu\rho}\;
\omega_{\nu\sigma} + \theta_{\mu\sigma}\;\omega_{\nu\rho} +
\theta_{\nu\rho}\;\omega_{\mu\sigma} +
\theta_{\nu\sigma}\;\omega_{\mu\rho} \right),\nonumber
\end{eqnarray}

\begin{eqnarray}
P^2_{\mu\nu,\;\rho\sigma} = \frac{1}{2} \left(
\theta_{\mu\rho}\;\theta_{\nu\sigma} +
\theta_{\mu\sigma}\;\theta_{\nu\rho} -
\theta_{\mu\nu}\;\theta_{\rho\sigma} \right), \nonumber
\end{eqnarray}

\begin{eqnarray}
P^0_{\mu\nu,\;\rho\sigma} = \frac{1}{2}
\theta_{\mu\nu}\;\theta_{\rho\sigma},\;\;{\overline
P}^{\;0}_{\mu\nu,\;\rho\sigma} =
\omega_{\mu\nu}\;\omega_{\rho\sigma}, \nonumber
\end{eqnarray}

\begin{eqnarray}
{\overline{\overline P}}^{\;0}_{\mu\nu,\;\rho\sigma} =
\theta_{\mu\nu}\;\omega_{\rho\sigma} +
\omega_{\mu\nu}\;\theta_{\rho\sigma},\nonumber
\end{eqnarray}

\noindent and $P$ is the operator associated with the linearized
Chern-Simons term, i.e.,

\begin{eqnarray}
 P_{\mu\nu,\;\rho\sigma} \equiv \frac{\Box
\partial^\lambda}{4}[\epsilon_{\mu\lambda\rho}\;\theta_{\nu\sigma}
+ \epsilon_{\mu\lambda\sigma}\;\theta_{\nu\rho} +
\epsilon_{\nu\lambda\rho}\;\theta_{\mu\sigma} +
\epsilon_{\nu\lambda\sigma}\;\theta_{\mu\rho}], \nonumber
\end{eqnarray}

\noindent since $ TP^1T = T{\overline P}^{\;0} T =
T{\overline{\overline P}}^{\;0} T =T P T =0$. The corresponding
multiplicative table is displayed in Table 2. The expansion of $O$
in the basis $\{P^1, P^2, P^0, {\overline P}^{\;0},
{\overline{\overline P}}^{\;0} , P \}$ is greatly facilitated if
use is made of the following tensorial identities:

\begin{eqnarray}
\frac{1}{2} (\eta_{\mu\rho} \eta_{\nu\sigma} + \eta_{\mu\sigma}
\eta_{\nu\rho} ) \equiv I_{\mu\nu,\;\rho\sigma} = [ P^1 + P^2 +
P^0 +{\overline P}^{\;0} ]_{\mu\nu,\;\rho\sigma}, \nonumber
\end{eqnarray}

\begin{eqnarray}
\eta_{\mu\nu} \eta_{\rho\sigma} = [2 P^0 +  {\overline P}^{\;0} +
{\overline{\overline P}}^{\;0}] _{\mu\nu,\;\rho\sigma},\;
\nonumber \frac{1}{\Box^2} (\partial_\mu \partial_\nu
\partial_\rho \partial_\sigma) = {{\overline
P}^{\;0}}_{\mu\nu,\;\rho\sigma}, \nonumber
\end{eqnarray}

\begin{eqnarray}
\frac{1}{\Box}(\eta_{\mu\rho} \partial_\nu \partial_\sigma +
\eta_{\mu\sigma}\partial_\nu \partial_\rho + \eta_{\nu\rho}
\partial_\mu \partial_\sigma + \eta_{\nu\sigma} \partial_\mu
\partial_\rho) = [2 P^1 + 4 {\overline P}^{\;0}] _{\mu\nu,
\;\rho\sigma}, \nonumber
\end{eqnarray}

\begin{eqnarray}
\frac{1}{\Box} (\eta_{\mu\nu} \partial_\rho \partial_\sigma +
\eta_{\rho\sigma} \partial_\mu \partial_\nu) =
[{\overline{\overline P}}^{\;0} + 2{\overline
P}^{\;0}]_{\mu\nu,\;\rho\sigma}. \nonumber
\end{eqnarray}

\vskip .5cm

\begin{center}
\begin{table}[h]
\caption{Multiplicative operator algebra fulfilled by $P^1$,
$P^2$, $P^0$, $\overline{P}^{\;0}$,
$\overline{\overline{P}}^{\;0}$ and  $P$. Here
${P^{\theta\omega}}_{\mu\nu\;,\;\rho\sigma}\equiv\theta_{\mu\nu}\omega_{\rho\sigma}$
and
${P^{\omega\theta}}_{\mu\nu\;,\;\rho\sigma}\equiv\omega_{\mu\nu}\theta_{\rho\sigma}$.
} \vskip .5cm
\begin{center}
\begin{tabular}{ccccccc}\hline
&$P^1$&$P^2$&$P^0$&${\overline P}^{\;0}$&${\overline{\overline P}
}^{\;0}$&$P$\\\hline
 $P^1$&$P^1$&0&0&0&0&0\\
$P^2$&0&$P^2$&0&0&0&$P$\\ $P^0$&0&0&$P^0$&0&$P^{\theta\omega}$&0\\
$\overline {P}^{\;0}$&0&0&0&${\overline
P}^{\;0}$&$P^{\omega\theta}$&0\\
 ${\overline{\overline P}
}^{\;0}$&0&0&$P^{\omega\theta}$&$P^{\theta\omega}$&$2(P^0+\overline
{P}^{\;0})$&0\\ $P$&0&$P$&0&0&0&$-{\Box}^3 P^2$\\\hline

\end{tabular}
\end{center}
\end{table}
\end{center}

\vskip .5cm

Expanding $O$ in the basis $\{P^1, P^2, P^0, {\overline P}^{\;0},
{\overline{\overline P}}^{\;0} , P \}$, we obtain $O = x_1 P^1 +
x_2 P^2 + x_0  P^0 + {\overline x}_{\,0} {\overline P}^{\;0} +
\overline{\overline {x}}_{\;0} {\overline{\overline P}}^{\;0} + p
P$. With the help of Table 2, we find that the propagator for MTMG
is given by

\begin{eqnarray}
O^{-1}_{\mathrm MTMG} &=& \frac{P^1}{x_1}  + \frac{x_2 P^2} {x_2^2
- p^2 k^6} + \frac{{\overline x}_{\,0} P^0}{x_0 {\overline x}_{0}
- 2 {\overline{\overline x}}_{0}^2} + \frac{x_0 {\overline
P}^{\;0}}{x_0 {\overline x}_{0} - 2 {\overline{\overline
x}}_{0}^2} \nonumber
\\ && -\; \frac {{\overline{\overline x}}_{\;0}
{\overline{\overline P}}^{\;0}}{x_0 {\overline x}_{\;0} - 2
{\overline{\overline x}}_{0}^2}  - \frac{pP}{ x^2_2 - p^2 k^6}.
\end{eqnarray}

Now, substituting eq. 6 into eq. 5, and taking  the identities,

\begin{eqnarray}
{P^2}_{\mu\nu,\;\rho\sigma} = \frac{1}{2} (\eta_{\mu\rho}
\eta_{\nu\sigma} + \eta_{\mu\sigma} \eta_{\nu\rho} ) - \frac{1}{2}
\eta_{\mu\nu} \eta_{\rho\sigma} -[ P ^1 + \frac{1}{2} {\overline
P}^{\;0} -\frac{1}{2} \overline{\overline{P}}^{\;0}
]_{\mu\nu,\;\rho\sigma}, \nonumber
\end{eqnarray}

\begin{eqnarray}
{P^0}_{\mu\nu,\;\rho\sigma} = \frac{1}{2} \eta_{\mu\nu}
\eta_{\rho\sigma} - \frac{1}{2} [{\overline P}^{\;0} +
{\overline{\overline P}}^{\;0}]_{\mu\nu,\;\rho\sigma}, \nonumber
\end{eqnarray}

\noindent  into account, yields

\begin{eqnarray}
SP_{\mathrm {MTMG}} =\left[ T^{\mu\nu} T_{\mu\nu} - \frac{1}{2}
T^2 \right] \frac{x_2 } {x_2^2 - p^2 k^6} + \frac{1}{2} T^2
\frac{{\overline x}_{\,0} }{x_0 {\overline x}_{0} - 2
{\overline{\overline x}}_{0}^2}.
\end{eqnarray}

\noindent We call attention to the fact that $f_{P^2} \equiv
\frac{x_2 } {x_2^2 - p^2 k^6}$ and $f_{P^0} \equiv
\frac{{\overline x}_{\,0} }{x_0 {\overline x}_{0} - 2
{\overline{\overline x}}_{0}^2}$ are, in this order, the
components $P^2$ and $P^0$ of $ O^{-1}_{\mathrm{MTMG}} $ in the
basis $\{P^1, P^2, P^0, {\overline P}^{\;0}, {\overline{\overline
P}}^{\;0} , P \}$.

The lemma that follows clears up the question of the sign of
$\;T^{\mu\nu} T_{\mu\nu} - \frac{1}{2} T^2$ at the physical poles;
it is also very useful for checking the presence of massless
spin-2 non-propagating excitations in the models we are analyzing.

\vskip .5cm

{\bf Lemma 2.} {\it If $m \geq 0$ is the mass of a generical
physical particle associated with the MTMG and $k$ is the
corresponding momentum exchanged, then $\left[T^{\mu\nu}
T_{\mu\nu} - \frac{1}{2} T^2 \right]_{k^2=m^2} >0$ and
$\;\left[T^{\mu\nu} T_{\mu\nu} -  T^2 \right]_{k^2=0} =0$.}

\vskip .5cm

{\bf Proof.} Using eq. 4, we can write the symmetric current
tensor as follows

$$T^{\mu\nu} = A k^\mu k^\nu + B{\tilde k}^\mu {\tilde k}^\nu + C
\varepsilon^\mu \varepsilon^\nu + Dk^{(\mu} {\tilde k}^{\nu)} + E
k^{(\mu} \varepsilon^{\nu)} + F{\tilde k}^{(\mu}
\varepsilon^{\nu)}. $$

The current conservation gives the following constraints for the
coefficients $A, B, D, E,$ and $F$:

\begin{equation}
Ak^2 + \frac{D}{2} \left( k^2_0 + \bf{k^2} \right) =0,
\end{equation}

\begin{equation}
B\left( k^2_0 + \bf{k^2} \right) + \frac{D}{2}k^2 =0,
\end{equation}

\begin{equation}
E k^2 + F\left( k^2_0 + \bf{k^2} \right) =0.
\end{equation}

From eqs. 8 and 9, we get $Ak^4 = B\left( k^2_0 + \bf{k^2}
\right)^2,$ while eq. 10 implies $E^2 > F^2$. On the other hand,
saturating the indices of $T^{\mu\nu}$ with momenta $k_{\mu}$, we
arrive at a consistent relation for the coefficients $A, B,$ and
$D$:
\begin{eqnarray}
Ak^4 + B\left( k^2_0 + \bf{k^2} \right)^2 + Dk^2 \left( k^2_0 +
\bf{k^2} \right) =0. \nonumber
\end{eqnarray}

After a lengthy but otherwise straightforward calculation using
the earlier equations, we obtain

\begin{eqnarray}
T^{\mu\nu} T_{\mu\nu} - \frac{1}{2} T^2=
\left[\frac{k^2(A-B)}{\sqrt{2}} - \frac{C}{\sqrt{2}} \right]^2 +
\frac{k^2}{2} (E^2 - F^2),\noindent
\end{eqnarray}

\begin{eqnarray}
T^{\mu\nu} T_{\mu\nu} - T^2 = k^2 \left[ \frac{1}{2} (E^2 - F^2)
-2C(A-B) \right]. \noindent
\end{eqnarray}
\noindent Therefore, $\left[ T^{\mu\nu} T_{\mu\nu} - \frac{1}{2}
T^2 \right]_{k^2= m^2} >0$ and $\left[ T^{\mu\nu} T_{\mu\nu} - T^2
\right]_{k^2= 0} =0.$

\vskip .5cm

We remark that $T^{\mu\nu} T_{\mu\nu} - \frac{1}{2} T^2$ is always
greater than zero for any physical particle; in addition,
$T^{\mu\nu} T_{\mu\nu} -  T^2$ is zero for  massless spin-2
non-propagating modes.

We are ready now to enunciate the algorithm for testing the
unitarity of the MTMG.

\vskip .5cm

{\bf Algorithm 2.} {\it Compute $SP_{\mathrm{MTMG}}$ using eq. 7
and then find the signs of the residues at each simple pole of
$SP_{\mathrm{MTMG}}$ with the help of the Lemma 2. If all the
signs are $\geq 0$ , the model is  unitary; however, if at least
one of the signs is negative, the system is non-unitary .}

\vskip .5cm

\noindent {\bf 3. CHECKING THE UNITARITY OF TMHDE AND TMHDG}

\vskip .5cm

We introduce here the two three-term systems we want to test the
unitarity, {\it i.e.}, TMHDE and TMHDG, and  afterwards we study
their unitarity.

\vskip .5cm

\noindent {\it 3.1  The models}

\vskip .5cm

The  Lagrangian for TMHDE is the sum of Maxwell, higher-derivative
(Podolsky and Schwed, 1948), gauge-fixing (Lorentz-gauge), and
Chern-Simons, terms, {\it i.e.},

\begin{equation}
{\mathcal L_{\mathrm{TMHDE}}} = - \frac{F_{\mu\nu} F^{\mu\nu}}{4}
+ \frac{l^2}{2} \partial_\nu F^{\mu\nu}\partial^\lambda
F_{\mu\lambda} - \frac{1}{2\lambda}(\partial_\nu A^\nu)^2 +
\frac{s}{2} \varepsilon_{\mu\nu\rho} A^\mu
\partial^\nu A^\rho.
\end{equation}

\noindent Here, $F_{\mu\nu} = \partial_\nu A_\mu - \partial_\mu
A_\nu$ is the usual electromagnetic tensor field, and $l$ is a
cutoff. The corresponding propagator is given by

\begin{equation}
O^{-1}_{\mathrm{TMHDE}} = \frac{l^2 k^4 + k^2}{(l ^2 k ^4 +k^2)^2
- s^2 k^2} \;\theta - \frac{\lambda}{k^2} \;\omega - \frac{s}{(l
^2 k ^4 +k^2)^2 - s^2 k^2}\;S.
\end{equation}

The Lagrangian related to TMHDG, in turn, is given by

\begin{eqnarray}
{\mathcal L}_{\mathrm{TMHDG}} &=& \sqrt{g} \left(
 - \frac{2R}{\kappa^2} +\frac{\alpha}{2} R^2 +\frac{ \beta}{2}
R_{\mu\nu}^2 \right) \nonumber \\ &&+ \frac{1}{\mu}
\epsilon^{\lambda\mu\nu} {\Gamma^{\rho}}_{\lambda\sigma} \left(
\partial_\mu {\Gamma^\sigma}_{\rho\nu} +\frac{2}{3}
{\Gamma^\sigma}_{\mu\beta}{\Gamma^\beta}_{\nu\rho}\right),
\end{eqnarray}

\noindent where $\alpha$ and $\beta$ are suitable constants with
dimension $L$. For the sake of simplicity, the  gauge-fixing term
was omitted. Linearizing  eq. 15 and adding to the result the
gauge-fixing term ${\mathcal{ L}}_{\mathrm {gf}} =
\frac{1}{2\lambda} ({h_{\mu\nu}}^{,\;\nu} -
\frac{1}{2}h_{,\;\mu})^2$ (de Donder gauge), we find  that the
propagator concerning TMHDG takes the form

\begin{eqnarray}
O^{-1}_{\mathrm{TMHDG}}&=& \frac{1}{\Box[-1
+b(\frac{3}{2}+4c)\Box]}\overline{\overline{P}}^{\;0} +
\frac{2\lambda}{k^2}P^1 +\;\frac{1}{\Box[-1
+b(\frac{3}{2}+4c)\Box]}P^0 \nonumber \\
&&+\;\frac{4M}{\Box[M^2b^2\Box^2+4(bM^2+1)\Box+4M^2]}P \nonumber
\\ &&\;\frac{2M^2(2+b\Box)}{\Box[M^2b^2\Box^2+ 4(bM^2+1)\Box+4M^2]}P^2\nonumber\\
&&+\;\left[
-\frac{4\lambda}{\Box}+\frac{2}{\Box[-1+b(\frac{3}{2}+4c)\Box]}\right]\overline
{P}^{\;0},
\end{eqnarray}
\noindent where $b\equiv\frac{\beta\kappa^2}{2}$, $c \equiv
\frac{\alpha}{\beta}$, and $M \equiv \frac{\mu}{\kappa^2}$.

\vskip .5cm

\noindent {\it 3.2 Testing the unitarity of TMHDE}

\vskip .5cm

The calculations  that are needed for  checking  the unitarity of
TMHDE are somewhat complicated because this model represents in
general three massive excitations.  Since the
 $\theta$-component of the propagator concerning TMHDE   can
 be written as $f_\theta=\frac{M^2(x-M^2)}{x^3 -2M^2x^2 + M^4x -
 M^4s^2}$, where $M\equiv \frac{1}{l}$,  we have to
 analyze the nature, as well as the signs, of the roots of the cubic equation $x^3
 + a_2x^2 + a_1x + a_0=0$, where $a_2 \equiv -2M^2, a_1 \equiv M^4,$ and $a_0
  \equiv -M^4 s^2$. Taking into account that we are only interested in
  those roots that are both real and unequal, we require
  $D<0$, where $D \equiv Q^3 + R^2$, with $Q$ and $R$ being, in
  this order, equal to $\frac{3a_1 - a^2_2}{9}$ and
  $\frac{9a_1a_2 -27a_0 -2a^3_2}{54}$, is the polynomial
  discriminant. Performing the computations we get $D=M^8s^2
  \left[ \frac{s^2}{4} - \frac{M^2}{27} \right]$, implying that
  only and if only $s^2 <  \frac{4M^2}{27}$ will the roots be real
  and distinct. Our next step is to verify whether or not these roots
  are positive. This can be accomplished by building the
  Routh-Hurwitz array (Uspensky, 1948), namely,

\begin{table}[h]
\begin{center}
\begin{tabular}{cc}
1&$M^4$\\ $-2M^2$&$-M^4s^2$\\ $M^2\left(M^2
-\frac{s^2}{2}\right)$&0\\ $-M^4s^2$&0
\end{tabular}
\end{center}
\end{table}

Noting that there are three signs changes in the first column of
the array above, we conclude that all the three  roots are
positive. In summary, if $s^2 < \frac{4m^2}{27}$, TMHDE   is a
model with acceptable values for the masses. Denoting these roots
as $x_1, x_2$, and $x_3$, and assuming without any loss of
generality that $x_1 > x_2 > x_3$, we get

\begin{eqnarray}
f_\theta  &=& \frac{M^2(x_1 -M^2)}{(x_1 -x_2)(x_1 - x_3)}
\frac{1}{x -x_1} + \frac{M^2(x_2 -M^2)}{(x_2 -x_1)(x_2 - x_3)}
\frac{1}{x -x_2} \nonumber \\ &&+ \frac{M^2(x_3 -M^2)}{(x_3
-x_1)(x_3 - x_2)} \frac{1}{x -x_3}. \nonumber
\end{eqnarray}

Hence, TMHDE with  will be unitary if the conditions $ x_1 -M^2<0,
\; x_2 -M^2>0,\;$ and $x_3 -M^2<0$ hold simultaneously. Obviously,
this will never occur, which allows us to conclude that TMHDE is
non-unitary.

Should we expect intuitively that TMHDE  faced unitary problems?
The answer is affirmative. In fact, setting $s=0$, for instance,
in its Lagrangian, we recover the Lagrangian for the usual
Podolsky electromagnetism which is non-unitary (Podolsky and
Schwed, 1948). Nonetheless, Podolsky-Chern-Simons (PCS) planar
electromagnetism with  $s^2 < \frac{4M^2}{27}$, despite being
haunted by ghosts, has normal massive modes. Note that the
existence of these well-behaved excitations is subordinated to the
condition
 $s< \frac{2M}{\sqrt{27}}$, which really encourages us to regard
 this system as an effective field model. We shall discuss their
 astonishing properties in Section 4.
\vskip .5cm

\noindent {\it 3.3 Testing the unitarity of TMHDG}

\vskip .5cm

The $SP$ concerning TMHDG can be written as

\begin{eqnarray}
SP_{\mathrm{TMHDG}}&=&\frac{M^2b}{2} \frac{
(T^{\mu\nu}T_{\mu\nu}-\frac{1}{2}T^2)}{ k^2 -M^2_1}
\frac{-1+\sqrt{1+2bM^2}}{\sqrt{1+2bM^2} \left[1 + bM^2
-\sqrt{1+2bM^2} \right] }\nonumber\\ &&+ \frac{M^2b}{2} \frac{
(T^{\mu\nu}T_{\mu\nu}-\frac{1}{2}T^2)}{ k^2 -M^2_2}
\frac{1+\sqrt{1+2bM^2}}{\sqrt{1+2bM^2} \left[1 + bM^2
+\sqrt{1-+bM^2} \right] }\nonumber\\ &&+\; -
\frac{T^{\mu\nu}T_{\mu\nu}-T^2}{k^2}-\frac{\frac{1}{2}T^2}{(k^2-m^2)},
\end{eqnarray}

\noindent where
\begin{eqnarray*}
M_1^2&\equiv&
\left(\frac{2}{b^2M^2}\right)[1+bM^2-\sqrt{1+2bM^2}],\nonumber\\
M_2^2&\equiv&
\left(\frac{2}{b^2M^2}\right)[1+bM^2+\sqrt{1+2bM^2}],\nonumber\\
m^2&\equiv&  - \frac{1}{b(3/2+4c )}.
\end{eqnarray*}

It is interesting to note that $M^2_1 \rightarrow M^2$, and $M^2_2
\rightarrow +\infty\;$, as $b \rightarrow 0$, implying that when
$\alpha,\beta \rightarrow 0$, eq. 17 reduces to

\begin{equation}
SP =  \left(T^{\mu\mu} T_{\mu\nu} - \frac{1}{2} T^2 \right)
\frac{1}{k^2-M^2} + \left(T^{\mu\mu} T_{\mu\nu} - T^2 \right)
\frac{1}{k^2},
\end{equation}

\noindent which is the expression for the $SP$ related to
Maxwell-Chern-Simons theory (MCS). Using eq. 18, we promptly
obtain  $$ {\mathrm{Res}} (SP)|_{\;k^2=M^2} > 0,\; {\mathrm{Res}}
(SP)|_{k^2=0} =0,$$

\noindent which means that MCS is unitary. Thence, we have
reobtained, in a trivial way, a well-known result (Deser {\it et
al.}, 1988a,b).

 We are now ready to analyze the excitations and mass counts
concerning TMHDG . To avoid needless repetitions, we restrict
ourselves to presenting a summary of the main results in Table 3.
The systems that do not appear in this table are tachyonic, {\it
i.e.}, unphysical. As intuitively expected, TMHDG is non-unitary.
Indeed, if the topologically massive term is removed, TMHDG
reduces to three-dimensinal higher-derivative gravity---an
effectively multimass model of the fourth-derivative order with
interesting properties of its own (Accioly {\it et al.},
2001a,b,c,)---which is non-unitary. Nonetheless, TMHDG is in
general non-tachyonic, which means that under circumstances it may
be viewed as an effective field model. We shall investigate, in
passing, the novel and amazing features of this  effective system
in Section 5.

\begin{table}[h]
\caption{Unitarity analysis of  topologically massive
higher-derivative gravity} \vskip 0.5cm
\begin{tiny}
\begin{center}
\begin{tabular}{ccccc}\hline
$b$ & $\frac{3}{2}+4c$ &\begin{tabular}{c} excitations and\\ mass
counts \end{tabular}& tachyons & unitarity
\\\hline $>0$&$<0$& \begin{tabular}{c} 2 massive\\
spin-2 normal particles\\ 1 massless spin-2\\ non-propagating
particle \\ 1 massive spin-0 ghost
\end{tabular}
&no one& \begin{tabular}{c}non-unitary
\end{tabular}\\\hline$\frac{-1}{2M^2}<b<0$&$>0$& \begin{tabular}{c} 1 massive\\
spin-2 normal particle\\ 1 massless spin-2\\ non-propagating
particle\\ 1 massive spin-2 ghost
\\ 1 massive spin-0 ghost
\end{tabular}
&no one& \begin{tabular}{c}non-unitary

\end{tabular}\\\hline
\end{tabular}
\end{center}
\end{tiny}
\end{table}

\vskip .5cm

\newpage

\noindent{\bf 4.  ATTRACTIVE INTERACTION BETWEEN EQUAL CHARGE
BOSONS IN THE FRAMEWORK OF MAXWELL-CHERN-SIMONS ELECTRODYNAMICS}

\vskip .5cm

In order to avoid extremely long calculations, we investigate here
Maxwell-Chern-Simons electrodynamics (MCSE) instead of
Podolsky-Chern-Simons electrodynamics (PCSE). Certainly, the two
models share similar characteristics. In other words, the exciting
features of PCSE are also present, {\it mutatis mutandis}, in
MCSE. Accordingly, let us analyze the interaction between equal
charge bosons in the context of the MCSE coupled to a
charged-scalar field. To do that we need to compute, first of all,
the effective non-relativistic potential for the interaction of
two charged-scalar bosons. Now, non-relativistic quantum mechanics
tells us that in the first Born approximation the cross section
for the scattering of two indistinguishable massive particles, in
the center-of-mass frame (CoM), is given by
$\frac{d\sigma}{d\Omega} = \left | \frac{m}{4\pi} \int e^{-i {\bf
p'} \cdot \; {\bf r}} V( r) e^{i{\bf p} \cdot\; {\bf r}}
d^2\;{\bf{r}} \right |^2,$ where ${\bf p}\; ({\bf p'})$ is the
initial (final) momentum of one of the particles in the CoM. In
terms of the transfer momentum, ${\bf k \equiv p' - p}$, it reads

\begin{eqnarray}
\frac{d\sigma}{d\Omega} = \left | \frac{m}{4\pi} \int V( r)
e^{i{\bf k} \cdot\; {\bf r}} d^{D-1}\;  {\bf{r}} \right |^2.
\end{eqnarray}

\indent On the other hand, from quantum field theory we know that
the cross section, in the CoM, for the scattering of two identical
massive scalars bosons by an electromagnetic field, can be written
a $\frac{d\sigma}{d\Omega} = \left | \frac{1}{16\pi E}\; {\cal
M}\right |^2,$ where $E$ is the initial energy of one of the
bosons and ${\cal M}$ is the Feynman amplitude for the process at
hand, which in the non-relativistic limit  (N.R.) reduces to

\begin{eqnarray}
\frac{d\sigma}{d\Omega} = \left | \frac{1}{16\pi m}\; {\cal
M}_{\mathrm{N.R.}} \right |^2.
\end{eqnarray}

\indent From eqs. 19 and 20 we come to the conclusion that the
expression that enables us to compute the effective
non-relativistic potential has the form

\begin{eqnarray}
V( r) = \frac{1}{4m^2} \frac{1}{(2\pi)^2} \int d^2\; {\bf k}\;
{\cal M}_{\mathrm{N.R.}}\; e^{-i {\bf k} \cdot \; {\bf r}},
\end{eqnarray}

\noindent which clearly shows how the potential from quantum
mechanics and the Feynman amplitude obtained via quantum field
theory are related to each other.

\indent Now, in the Lorentz gauge the MCSE coupled to a
charged-scalar field is described by the Lagrangian

\begin{eqnarray}
{\cal L} &=& -\frac{1}{4} F_{\mu\nu} F^{\mu\nu} +\frac{s}{2}
\varepsilon_{\mu\nu\rho} A^\mu \partial ^\nu A^\rho
-\frac{1}{2\lambda} (\partial_\nu A^\nu)^2 \nonumber\\ &&+ (D_\mu
\phi)^* D^\mu \phi - m^2 \phi^* \phi,
\end{eqnarray}

\noindent where $D_\mu \equiv \partial_\mu + iqA_\mu$. Therefore,
the interaction Lagrangian to order $Q$ for the process $S + S
\longrightarrow S + S$, where $S$ denotes a spinless boson of mass
$m$ and charge $Q$, is ${\cal L}_{int} = iQ A^\mu \left( \phi
\partial_\mu \phi^* - \phi^*
\partial_\mu \phi \right),$ implying that the elementary vertice
is given by $$\Gamma^\mu_\phi (p, p') = -Q(p +p')^\mu,$$

\noindent where $p$ $(p')$ is the momentum of the incoming
(outgoing) scalar boson. As a consequence, the Feynman amplitude
for the interaction of two charged spinless bosons of equal mass
is
\begin{equation}
{\cal M} =\Gamma^\mu_\phi (p, p') O^{-1}_{\mu\nu} \Gamma^\nu_\phi
(q, q')
\end{equation}

\noindent where

\begin{eqnarray}
O^{-1} =- \frac{ \theta}{k^2 - s^2} -\frac{\lambda \omega}{k^2} -
\frac{s S}{k^4- s^2k^2} . \nonumber
\end{eqnarray}

In the non-relativistic limit, the Feynman  amplitude for the
process under consideration assumes the form

\begin{eqnarray}
{\cal M}_{NR} =  \left[ \frac{4Q^2m^2}{ {\bf k}^2  + s^2} +
\frac{8ismQ^2 \; {\bf k} \wedge {\bf P} }{{\bf k}^2 ( {\bf k}^2 +
s^2)} \right], \nonumber
\end{eqnarray}

\noindent where ${\bf P} \equiv \frac{1}{2} ({\bf p} -{\bf q})$ is
the relative momentum of the incoming charged-scalar bosons in the
CoM.

It follows that the effective non-relativistic potential is given
by

\begin{eqnarray}
V(r) = -\frac{Q^2}{m\pi s} \left[ \frac{1}{r^2} - \frac{s K_1
(sr)}{r} \right] {\bf L} + \frac{Q^2}{2\pi s} K_0 (sr),
\end{eqnarray}

\noindent where $ {\bf L} \equiv {\bf r} \wedge {\bf P}$ is the
orbital angular momentum, and $K$ is the modified  Bessel
function.  Let us then investigate whether or not this potential
can bind a pair of identical charged-scalar bosons. In this case,
the corresponding time-independent Schr\"odinger equation can be
written as

\begin{eqnarray}
{\cal{H}}_l {\cal{R}}_{nl} &=& -\frac{1}{m} \left( \frac{d^2}{d
r^2} {\cal{R}}_{nl} + \frac{1}{r} \frac{d}{d r} {\cal{R}}_{nl}
\right) + V^{eff}_l {\cal{R}}_{nl} \nonumber \\ &=& E_{nl} {\cal
R}_{nl},
\end{eqnarray}

\begin{eqnarray}
V^{eff}_l &\equiv& \frac{l^2}{m r^2} + V(r) \nonumber \\ &=&
\frac{l^2}{m r^2} -\frac{Q^2}{m\pi s} \left[ \frac{1}{r^2} -
\frac{s K_1 (sr)}{r} \right] {\bf L} + \frac{Q^2}{2\pi s} K_0 (sr)
, \nonumber
\end{eqnarray}

\noindent where ${\cal R}_{nl}$ is the $n$th normalizable
eigenfunction of the radial Hamiltonian ${\cal H}_l$ whose
corresponding eigenvalue is $E_{nl}$ and $V^{eff}_l$ is the $l$th
partial wave effective potential. Note that $V^{eff}_l$ behaves as
$\frac{l^2}{mr^2}$ at the origin and as $\frac{l}{m} \left[\; l -
\frac{Q^2 s}{\pi s} \right] \frac{1}{r}$ asymptotically. On the
other hand,

\begin{eqnarray}
\frac{d}{d r} V^{eff}_l = -\frac{2l}{m} \left[l -\frac{Q^2}{\pi s}
\right] \frac{1}{r^3} - \frac{Q^2 sl}{m \pi} \frac{1}{r} K_0(sr) -
\left[ \frac{Q^2 2l}{m\pi r^2} + \frac{Q^2 s}{2\pi}
\right]K_1(sr)\nonumber
\end{eqnarray}

\noindent Assuming, without any loss of generality, that $l>0$, it
is trivial to see that, if $l> \frac{Q^2}{\pi s}$, the potential
is strictly decreasing, which precludes the existence of bound
states. The remaining possibility is $l< \frac{Q^2}{\pi s}$. In
this interval  $V^{eff}_l$ approaches $+ \infty$ at the origin and
$0^{-}$ for $r \rightarrow +\infty$, which is indicative of a
local minimum. Consequently, the existence of charged- scalar-
boson---charged-scalar-boson bound states is subordinated to the
condition $0< l < \frac{Q^2}{\pi s}$. In terms of the
dimensionless parameters $y \equiv s r,\; \alpha \equiv
\frac{Q^2}{\pi s},\; \beta \equiv \frac{m}{s},$ and $\; \tilde
{E}_{n l} \equiv \frac{mE_{n l}}{s^2}$, eq. 25 reads

\begin{eqnarray}
\left[\frac{d^2}{d y^2}+  \frac{1}{y} \frac{d}{d y} \right]
{\mathcal {R}}_{nl} + \left[\tilde {E}_{n l} - \tilde{V}^{eff}_{l}
\right] {\mathcal {R}}_{n l}=0,
\end{eqnarray}

with

\begin{eqnarray}
\tilde{V}^{eff}_{l} \equiv -\frac{l (\alpha - l)}{y^2}
+\frac{\alpha \beta}{2}  K_0( y) -\frac{\alpha l}{y} K_1(
y).\nonumber
\end{eqnarray}

\noindent Of course, eq. 26 cannot be solved analytically;
nevertheless, it can be solved numerically. To accomplish this, we
rewrite the radial function as ${\cal R}_{nl} \equiv \frac{u_{n
l}}{\sqrt{y}} $. As a consequence, eq.
 26 takes the form

\begin{eqnarray}
\left[\frac{d^2}{d y^2} + \frac{1}{4 y^2} \right]u_{n l} + \left[
\tilde{E}_{n l} - \tilde{V}^{eff}_{l}\right]u_{n l}.
\end{eqnarray}

\noindent Using the Numerov algorithm (Numerov, 1924), we have
solved eq. 27 numerically for several values of the parameters
$\alpha,\beta,$ and $l$. In Fig.
 1 we present our numerical
results for the potential in the specific case of $l= 6$. The
corresponding ground-state energy is $-1.68 \times 10^{-8}$ MeV.
The graphic shown in Fig. 1 exhibits the generic features of the
potential , although it has been composed using particular values
of the parameters $\alpha,\beta,$ and $l$.

\begin{figure}
\begin{center}
\includegraphics[scale=.8]{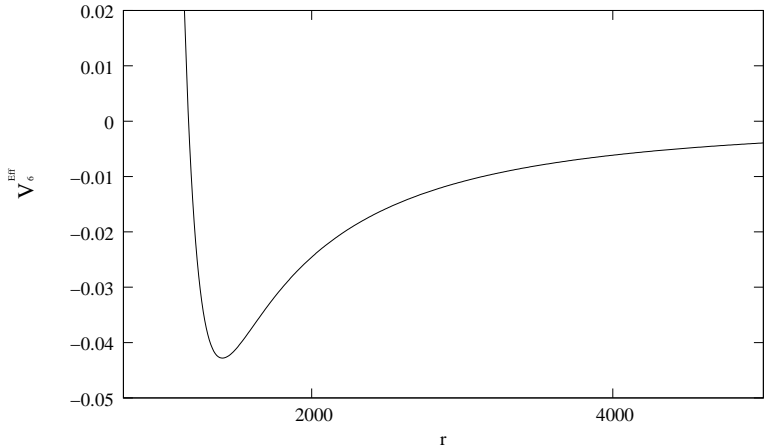}
\end{center}
\caption{Attractive effective non-relativistic potential
corresponding to the eigenvalue $l =6$ . Here $[V^{eff}_6] =$ eV,
$[r] ={\mathrm{ MeV^{-1}}}$, $\alpha =7.6$, and $\beta =7000$. }
\end{figure}

In conclusion we may say that since ``Cooper pairs"   exist in the
framework of MCSE, they also exist, as a consequence, in the
context of PCSE. A detailed study of the potential, as well as the
eigenvalue structure, for the PCSE coupled with a charged-scalar
field, will be published elsewhere (Accioly and Dias, 2004a)

\vskip .5cm

\noindent {\bf 5. GRAVITY, ANTIGRAVITY, AND GRAVITATIONAL
SHIELDING IN THE CONTEXT OF THREE-DIMENSIONAL GENERAL RELATIVITY
WITH HIGHER-DERIVATIVES}

\vskip .5cm

For reasons similar to those discussed in Section 4, we consider
here the astonishing features of higher-derivative gravity instead
of TMHDG. Let us then compute the effective non-relativistic
potential for the interaction of two identical massive bosons of
zero spin via a graviton exchange. The expression for the
potential is

\begin{eqnarray}
V( r) = \frac{1}{4m^2} \frac{1}{(2\pi)^2} \int d^2\; {\bf k}\;
{\cal M}_{\mathrm{N.R.}}\; e^{-i {\bf k} \cdot \; {\bf r}},
\end{eqnarray}

\noindent where $m$ is the mass of one of the bosons. Now, the
interaction Lagrangian for the process we are analyzing is

\begin{eqnarray}
{\cal L}_{\mathrm{int}} = -\frac{\kappa h^{\mu\nu}}{2} \left[
\partial_\mu \phi \partial_\nu \phi - \frac{1}{2} \eta_{\mu\nu}
\left( \partial_\alpha \phi \partial^\alpha \phi - m^2 \phi^2
\right)\right], \nonumber
\end{eqnarray}

\noindent implying that the elementary vertice can be written as

\begin{eqnarray}
\Gamma^\phi_{\mu\nu} (p, p') = \frac{1}{2} \kappa\left[ p_\mu
p'_\nu + p_\nu p'_\mu - \eta_{\mu\nu} \left(p.p' + m^2 \right)
\right],
\end{eqnarray}

\noindent where the momenta are supposed to be incoming. The
expression for the non-relativistic Feynman amplitude is, in turn,
given by

\begin{eqnarray}
{\cal M}_{\mathrm{N.R.}} = -\frac{1}{2} \frac{\kappa^2 m^4
m^2_1}{{\bf k^2} ({\bf k^2} + m^2_1)} + \frac{1}{2} \frac{\kappa^2
 m^4 m^2_0}{{\bf k^2} ({\bf k^2} + m^2_0)},
\end{eqnarray}

\noindent where $m^2_0 \equiv \frac{1}{\kappa^2 [\frac{3}{4} \beta
 + 2\alpha]}$ and $m^2_1 \equiv - \frac{4}{\kappa^2 \beta}$ are
 supposed to be positive in order to avoid the presence of
 tachyons in the dynamical field. Performing the appropriate
 integrations using eqs. 28 and 30, we obtain the effective
 non-relativistic potential, namely,

\begin{eqnarray}
V(r) = 2Gm^2 \left[ K_0 (m_1 r) - K_0 (m_0 r) \right].
\end{eqnarray}

\noindent Note that $V(r)$ behaves as $2Gm^2 \ln
(\frac{m_0}{m_1})$ at the origin and as $$2Gm^2 \left[
\sqrt{\frac{\pi}{2m_1 r}}\; e^{- m_1 r} - \sqrt{ \frac{\pi}{2m_0
r}}\; e^{- m_0 r} \right]$$

\noindent asymptotically. Note that this potential is extremely
well-behaved: it is finite at the origin and zero at infinity. On
the other hand, the derivative of the potential with respect to
$r$ is given by

\begin{eqnarray}
\frac{dV}{dr} = 2Gm^2 \left[ -m_1 K_1 (m_1 r) + m_0 K_1 (m_0 r)
\right], \nonumber
\end{eqnarray}

\noindent implying that it is everywhere attractive if $m_0
> m_1$, is repulsive if $ m_1 > m_0$, and vanishes if $ m_1 =
m_0$. If we appeal to the usual tools of Einstein's geometrical
theory, we arrive at the same conclusions. In fact, in the weak
field approximation the gravitational acceleration , $\gamma^l =
\frac{dv^l}{dt}$, of a slowly moving  particle is given by
$\gamma^l = -\kappa \left[\; \partial_t h^l_0 - \frac{1}{2}
\partial^l h_{00} \right]$, which for time-independent fields
reduces to $\gamma^l = \frac{\kappa}{2} \partial^l h_{00}$. Now,
taking into account that $h_{00} = \frac{2V}{ m \kappa}$, we
obtain

 $$ \gamma^l = 2 m G \frac{x^l}{r} \left[ m_0
K_1(r m_0) - m_1 K_1 \left(m_1 r  \right) \right].$$

\noindent Therefore, the gravitational force exerted on the
particle  , $$F^l = 2 G m^2 \frac{x^l}{r}   \left[ m_0 K_1(r m_0)
- m_1 K_1 \left(m_1 r  \right) \right],$$

\noindent  is everywhere attractive if $m_0
> m_1$, is repulsive if $ m_1 > m_0$ (antigravity), and vanishes
if $m_1 = m_0$ (gravitational shielding). It is remarkable that
this force does not exist in general relativity. It is peculiar to
 both  higher-derivative gravity and
 TMHDG (Accioly and Dias, 2005).

In Fig. 2  it is shown a schematic picture of the effective
non-relativistic potential for the three situations described
above, {\it i.e.}, $m_0
> m_1$, $ m_1 > m_0$, and $m_1 = m_0$.

\begin{figure}
\begin{center}
\includegraphics[scale=1]{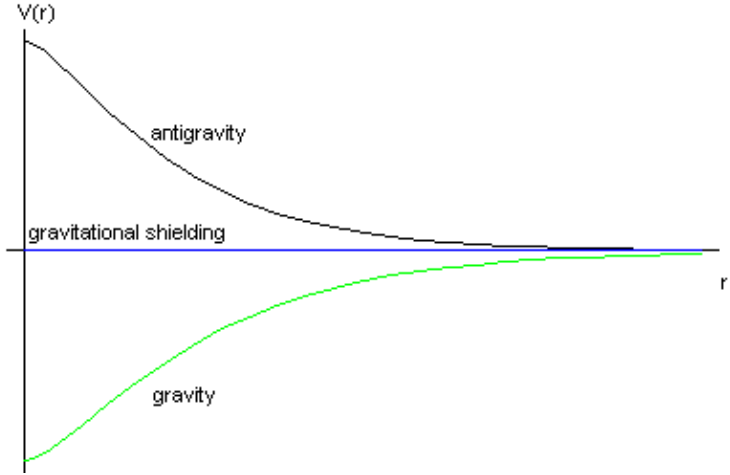}
\end{center}
\caption{ Gravity, antigravity and gravitational shielding in the
framework of three-dimensional Einstein's gravity with
higher-derivatives. }
\end{figure}

\vskip .5cm

\noindent {\bf 6. DISCUSSIONS AND COMMENTS}

\vskip .5cm

According to a somewhat obscure  unitarity lore it is expected
that the operation of augmenting a non-topological massive gravity
model through the topological term would transform the non-unitary
systems into unitary ones and preserve the unitarity of the
originally unitary models. This false idea is, perhaps,
responsible for the claims in the literature concerning the
pseudo-unitarity of both topologically massive Fierz-Pauli gravity
(TMFPG) (Pinheiro {\it et al.}, 1997a,b) and TMHDG (Pinheiro {\it
et al.}, 1997c). The authors of these works wrongly state that
these models are unitary. As far as TMPFG is concerned, it was
shown recently that this system with the Einstein's term with the
``wrong sign" is forbidden, while the model with the usual sign
has acceptable mass ranges but faces ghosts problems (Deser and
Tekin, 2002). On the other hand, the non-unitarity problem of
TMHDG was recently rehearsed (Accioly, 2003; Accioly, 2004) and
carefully tackled (Accioly and Dias, 2004b). In truth, we may say
that we will never be ale to construct an unitary, massive,
topologically massive, gravitational model. Indeed,  the fancy way
Einstein-Chern-Simons theory is built, {\it i.e.}, with the
Einstein's term with the opposite sign, precludes the existence of
ghost-free, massive, topologically massive, gravitational models
(Accioly and Dias, 2005). It is worth mentioning that these
idiosyncrasies do not occur in the framework of of massive,
topologically massive, electromagnetic models because the Maxwell
sign's term concerning Maxwell-Chern-Simons theory is the same as
that of the usual Maxwell's theory.

Nonetheless, the massive topologically massive models with higher
derivatives may be utilized  under certain circumstance  as
effective field models, {\it i.e}, as low-energy approximations to
more fundamental theories that, quoting Weinberg ( Weinberg,
1995), ``may not be field theories at all". The physics associated
with these models is not only intriguing, but also fascinating.
Certainly it deserves to be much better known.

\vskip .5cm

\noindent {\bf ACKNOWLEDGMENTS}

\vskip .5cm

We are very grateful to Prof. S. Deser for calling our attention
to the article ``Massive, Topologically Massive, Models" (Deser
and Tekin, 2002). A. Accioly thanks CNPq-Brazil for partial
support, while M. Dias thanks CAPES-Brazil for full support.

\vskip .5cm

\noindent {\bf REFERENCES}

\vskip .5cm

\noindent {\small Accioly, A., Azeredo, A., and Mukai, H. (2001a).
 {\it Physics Letters A} {\bf 279}, 169.}

\vskip .2cm

\noindent {\small Accioly, A., Mukai, H., and Azeredo, A. (2001b).
 {\it Classical and Quantum Gravity} {\bf 18}, L31.}

\vskip .2cm

\noindent {\small Accioly, A., Mukai, H., and Azeredo, A. (2001c).
 {\it Modern Physics Letters A} {\bf 16}, 1449.}

\vskip .2cm

\noindent {\small Accioly, A. (2003).
 {\it Physical Review D} {\bf 67}, 127502.}

\vskip .2cm

\noindent {\small Accioly, A. (2004).
 {\it Nuclear Physics B (Proceedings Supplements)} {\bf 127}, 100.}

\vskip .2cm

\noindent {\small Accioly, A., and Dias, M. (2004a),
 {\it Physical Review D} {\bf 70}, 107705.}
 \vskip .2cm

\noindent {\small Accioly, A., and Dias, M. (2004b),
 {\it Modern Physics Letters A} {\bf 19}, 817.}

\noindent {\small Accioly, A., and Dias, M. (2005),
 {\it International Journal of Modern Physics A,} in press.

\vskip .2cm

\noindent {\small Antoniadis, I., and Tomboulis, E. (1986).
 {\it Physical Review D} {\bf 33}, 2756.}

\vskip .2cm

\noindent {\small Deser, S., Jackiw, R., and Templeton, S.
(1982a).
 {\it Physical Review Letters} {\bf 48}, 475.}

\vskip .2cm

\noindent {\small Deser, S., Jackiw, R., and Templeton, S.
(1982b).
 {\it Annals of Physics} {\bf 140}, 372.}

\vskip .2cm

\noindent {\small Deser, S., and Tekin, G. (2002).
 {\it Classical and Quantum Gravity} {\bf 19}, L97.}

\vskip .2cm

\noindent {\small   Nieuwenhuizen, P. (1973).
 {\it Nuclear Physics B } {\bf 60}, 478.}

\vskip .2cm

\noindent {\small   Numerov, B. (1924).
 {\it Monthly Notes of the Royal Astronomical Society } {\bf 84}, 592.}

\vskip .2cm

\noindent {\small Pinheiro, C, Pires, G., and  Tomimura, N.
(1996a). {\it Il Nuovo Cimento B} {\bf 111} 1023.}

\vskip .2cm

\noindent {\small  Pinheiro, C.,  Pires, G., and  Rabelo de
Carvalho, F. (1997b). {\it Brazilian Journal of  Physics} {\bf 27}
14.}

\vskip .5cm

{\noindent {\small  Pinheiro, C., Pires, G., and  Sasaki, C.
(1997c). {\it General Relativity and  Gravitation} {\bf 29}  409.}

\vskip .5cm

{\noindent {\small  Podolsky, B., and  Schwed, M. (1948). {\it
Reviews of Modern Physics} {\bf 20}  40.}

\vskip .5cm

{\noindent {\small  Rivers, R. (1964). {\it Il Nuovo Cimento} {\bf
34} 387.}

\vskip .2cm

\noindent {\small Stelle, K. (2003).
 {\it Physical Review D} {\bf 16}, 953.}

\vskip .2cm

\noindent {\small Uspensky, J. (1948).
 {\it Theory of Equations}, McGraw-Hill, New York. }

\vskip .2cm

\noindent {\small Weinberg., S. (1995).
 {\it The Quantum Theory of Fields}, Volume I, Cambridge University Press, Cambridge. }

\end{document}